\documentclass[3p,times,twocolumn]{elsarticle}

\usepackage{ecrc}

\volume{260}

\firstpage{111}

\journalname{Nuclear Physics B Proceedings Supplement}

\runauth{P. Masjuan}

\jid{nuphbp}

\jnltitlelogo{Nuclear Physics B Proceedings Supplement}

\usepackage{amssymb}

\usepackage[figuresright]{rotating}

\begin{document}
 
\begin{frontmatter}


\title{Overview of the hadronic light-by-light contribution to the muon $(g-2)$}
\author{Pere Masjuan}
\ead{masjuan@kph.uni-mainz.de}
\address{PRISMA Cluster of Excellence and Institut f\"ur Kernphysik, Johannes Gutenberg-Universit\"at Mainz, D-55099 Mainz, Germany}
\fntext[label3]{eprint: MITP/14-088}


\title{Overview of the hadronic light-by-light contribution to the muon $(g-2)$}


\author{}

\address{}

\begin{abstract}
In this talk I review the recent progress on the numerical evaluation of the Hadronic Light-by-Light contribution to the anomalous magnetic moment of the muon and I discuss the role of experimental data on the accuracy of its determination.
\end{abstract}


\end{frontmatter}


\section{Introduction}
\label{intro}

The anomalous magnetic moment of the muon $(g-2)_{\mu}$ is one of the most accurately measured quantities in particle physics, and as such is a very promising signal of new physics if a deviation from its prediction in the Standard Model is found. 

The present experimental value for $a_{\mu}=(g-2)_{\mu}/2$, is given by $a_{\mu}^{\mathrm{EXP}}=116 592 09.1(6.3)\times10^{-10}$, as an average of $a_{\mu^+}=116 592 04(7.8)\times10^{-10}$ and $a_{\mu^-}=116 592 15(8.5)\times10^{-10}$ \cite{Bennett:2004pv,PDG2014}. Since statistical errors are the largest source of uncertainties, a proposal to measure it again to a precision of $1.6 \times 10^{-10}$ has recently been submitted to FNAL \cite{Carey:2009zzb} and JPARC~\cite{JPARC}, using different experimental techniques.

At the level of the experimental accuracy, the QED contributions has been completed up to the fifth order ${\cal O}(\alpha_{em}^5)$, giving the QED contribution $11658471.895(8)\times10^{-10}$ \cite{Aoyama:2012wk}, using the Rydberg constant and the ratio $m_{Rb}/m_e$ as inputs~\cite{PDG2014}. Also electroweak (EW) and hadronic contributions are necessary. The latter represents the main uncertainty in the Standard Model in terms of the hadronic vacuum polarization (HVP) and the hadronic light-by-light scattering (HLBL). The present estimates for QED, HVP, HLBL, and EW corrections are collected in Table~\ref{SMcont}.

\begin{table}[htbp]
\begin{center}
\begin{tabular}{ccc}
Contribution  &  Result in $10^{-10}$ units & Ref.\\[5pt]
\hline
QED (leptons) & $11658471.895\pm 0.008$ & \cite{Aoyama:2012wk}\\
HVP (leading order) & $694.9\pm4.3$ & \cite{Hagiwara:2011af}\\
HVP$^{\mathrm{NLO+NNLO}}$ & $-8.6\pm0.1$& \cite{Hagiwara:2011af,Kurz:2014wya}\\
HLBL & $11.6 \pm4.0$ & \cite{Jegerlehner:2009ry}\\
EW & $15.36\pm0.010$ &\cite{Gnendiger:2013pva}\\
\hline
Total & $11659185.2\pm5.9$\\
\hline
\end{tabular}
\end{center}
\caption{Standard Model contributions to $(g-2)_{\mu}$.}
\label{SMcont}
\end{table}%

For the HLBL, two reference numbers can be found in the literature: $a_{\mu}^{{\mathrm{HLBL}}} = (11.6\pm4.0)\times10^{-10}$~\cite{Jegerlehner:2009ry} but also $(10.5\pm2.5)\times10^{-10}$~\cite{Prades:2009tw}. They imply a discrepancy $\Delta a_{\mu} = a_{\mu}^{\mathrm{EXP}} - a_{\mu}^{\mathrm{SM}}$ of about  $2.7\sigma$ and $2.9\sigma$, respectively. The overall HLBL contribution is of the order of the present experimental error determination, which means that if, as an amusement, we discard it, $\Delta a_{\mu}$ increases by $2\sigma$ (the size of the HLBL). The striking situation then comes when the foreseen experiments (precision of $1.6\times 10^{-10}$) would imply the HLBL being a $5\sigma$ effect. On the light of such numbers we really need to understand the HLBL values and its errors.

The progress on the field is captured in at least three recent dedicated workshops on $(g-2)_{\mu}$~\cite{Czyz:2013zga,Benayoun:2014tra}, being the summary talk of the topical parallel session on \emph{photon-photon physics and its implications for the muon's $(g-2)$} of the $10^{th}$ EINN2013 (http://cyprusconferences.org/einn2013/program.php) a good overview.

\begin{table*}[htdp]
\caption{The HLBL and its different contributions from different references and methods, representing the progress on the field and the variety of approaches considered. $\dag$ indicates used from a previous calculation.}
\begin{center}
\begin{tabular}{llccccl}
Authors \& Refs. & HLBL & $\pi,K$ loop & PS ex. & higher spin ex. & quark loop & method and year\\
\hline
BPP~\cite{Bijnens:1995cc} &$+83(32)$ & $-19(13)$ &  $+85(13)$ &  $-4(3)$ &  $+21(3)$ & \textrm{ENJL, '95\, '96\, '02}\\  
HKS~\cite{Hayakawa:1995ps} &$+90(15)$ & $-5(8)$ &  $+83(6)$ &  $+1.7(1.7)$ &  $+10(11)$& \textrm{LHS, '95\, '96\, '02}  \\  
KN~\cite{Knecht:2001qf} &$+80(40)$ & & $+83(12)$ &  & &\textrm{Large $N_c$+$\chi$PT, '02}\\
MV~\cite{Melnikov:2003xd} &$+136(25)$ & $0(10)$ &  $+114(10)$ &  $+22(5)$ &  $0$&\textrm{Large $N_c$+$\chi$PT, '04} \\    
JN~\cite{Jegerlehner:2009ry} &$+116(40)$ & $-19(13)\dag$ &  $+99(16)$ &  $+15(7)$ &  $+21(3)\dag$&\textrm{Large $N_c$+$\chi$PT, '09}\\  
PdRV~\cite{Prades:2009tw} &$+105(26)$ & $-19(19)$ &  $+114(13)$ &  $+8(12)$ &  $0$&\textrm{Average, '09}\\  
HK~\cite{Hong:2009zw} &$+107$ &  & $+107$ &  && \textrm{Hologr. QCD, '09} \\  
DRZ~\cite{Dorokhov:2011zf} &$+59(9)$ &  &$+59(9)$ &  && \textrm{Non-local q.m., '11} \\  
EMS~\cite{Masjuan:2012wy,Masjuan:2012qn,Escribano:2013kba} &$+107(17)$ &  $-19(13)\dag$ &  $+90(7)$ &   $+15(7)\dag$ &  $+21(3)\dag$ &\textrm{Pad\'e-data,'12\, '13}\\ 
EMS~\cite{Masjuan:2012sk,Escribano:2013kba} &$+105(16)$ &  $-19(13)\dag$ &  $+88(4)$ &   $+15(7)\dag$ &  $+21(3)\dag$ &\textrm{Large $N_c$ , '13} \\   
GLCR~\cite{Roig:2014uja} &$+118(20)$ & $-19(13)\dag$ &  $+105(5)$ &   $+15(7)\dag$ &  $+21(3)\dag$& \textrm{R$\chi$T, '14} \\  
\end{tabular}
\end{center}
\label{T2}
\end{table*}%

\section{Dissection of the HLBL and potential issues}

The HLBL cannot be directly related to any measurable cross section and requires knowledge of QCD at all energy scales. Since this is not known yet, one needs to rely on hadronic models to compute it. Such models introduce some systematic errors which are difficult to quantify. Using the large-$N_c$ and the chiral counting, de Rafael proposed~\cite{deRafael:1993za} to split the HLBL into a set of different contributions: pseudo-scalar exchange (dominant~\cite{Jegerlehner:2009ry,Prades:2009tw}), charged pion and kaon loops, quark loop, and higher-spin exchanges (see Table~\ref{T2}). The large-$N_c$ approach however has at least two shortcomings: firstly, it is difficult to use experimental data in a large-$N_c$ world. Secondly, calculations carried out in the large-$N_c$ limit demand an infinite set of resonances. As such sum is not known, one truncates the spectral function in a resonance saturation scheme, the Minimal Hadronic Approximation (MHA)~\cite{Peris:1998nj}. The resonance masses used in each calculation are then taken as the physical ones from PDG~\cite{PDG2014} instead of the corresponding masses in the large-$N_c$ limit. Both problems might lead to large systematic errors not included so far~\cite{Masjuan:2012wy,Masjuan:2007ay}.

In Table~\ref{T2}, I collect few of the main approaches for the HLBL. Among them, the Jegerlehner and Nyffeler review~\cite{Jegerlehner:2009ry} together with the \emph{Glasgow consensus} written by Prades, de Rafael, and Vainshtein~\cite{Prades:2009tw} represent, in my opinion, the two reference numbers. They agree well since they only differ by few subtleties. For the main contribution, the pseudoscalar exchange, on needs a model for the pseudoscalar Transition Form Factor (TFF). They both used the model from Knecht and Nyffeler~\cite{Knecht:2001qf} based on MHA, but differ on how to implement the high-energy QCD constrains coming from the VVA Green's function. In practice, this translates on wether the piece contains a pion pole or a pion exchange. The former would imply that the exchange of heavier pseudoscalar resonances (6th column in Table~\ref{T2}) is effectively included in~\cite{Melnikov:2003xd}, while the latter demands its inclusion. The other difference is wether the errors are summed linearly~\cite{Jegerlehner:2009ry} or in quadrature~\cite{Prades:2009tw}. 

Neither of both approaches contain systematic errors from chiral and large-$N_c$ limits~\cite{Masjuan:2012wy,Masjuan:2012qn,Masjuan:2007ay}. In the large $N_c$, the MHA should be understood from the mathematical theory of Pad\'e approximants (PA) to meromorphic functions~\cite{Masjuan:2007ay}. Obeying the rules from this mathematical framework, one can compute the desired quantities in a model-independent way and even be able to ascribe a systematic error to the approach \cite{Masjuan:2007ay}. Interestingly~\cite{Queralt:2010sv}, given the low-energy expansion of the TFF used here, its PA sequence converges much faster than the MHA, especially when the QCD behavior is imposed. In principle, one knows then how to incorporate large-$N_c$ systematics, but that task should still be done. On top, with new experimental data, the inputs for models used should be updated. Beyond that fact, it is common to factorize the TFF as a product of a single virtual form factor, effect never considered so far.

All in all, even though the QCD features for the HLBL are well understood~\cite{Jegerlehner:2009ry,Prades:2009tw}, the details of the particular calculations are important. Considering the drawback drawn here, I think we need more calculations, closer to experimental data if possible. 

\section{The role of experimental data on the HLBL}

Before going into detail, allow me an excursus on a recent lattice QCD simulation. Blum \textit{et al.}~\cite{Blum:2014oka} proposed a method for simulating the HLBL in a lattice QCD+QED. They studied a non-perturbative treatment of QED which later on was checked against the perturbative simulation. With that spirit, they considered that a QCD+QED simulation could deal with the non-perturbative effects of QCD for the HLBL. Unphysical quark and muon masses are used~\cite{Blum:2014oka}, and only the single quark-loop diagram is simulated, but still a lattice signal is obtained. Due to the finiteness of the volume, the simulation is not yet at zero momentum as the physical counterpart, but at $2\pi/L$ ($L$ the lattice spacing). The next step will be, then, to go for physical values, consider larger volumes and latter on control the extrapolation to the desired zero momentum point.

Going back to our data driven approaches, one of the recent progress on the field is the consideration of dispersion relations (DR) for calculating the HLBL. As example, Colangelo \textit{et al.}~\cite{Colangelo:2014dfa} considered a DR for the four-point tensor that leads to the HLBL. After decomposing it in terms of helicity amplitudes, and consider independently the contributions from scalar QED, one $\pi^0$ and $2 \pi$, they showed that the DR would need the $\pi^0$ TFF and the $\gamma^* \gamma^* \to \pi \pi$ as input, only. One expects good precision in the momentum region where such DR is formulated. In this framework, the TFF describes a pion-pole, no intermediate states are considered. Off-shell effects would be included in subtraction constants. For a reliable numerical evaluation the DR should extended up to infinity. Since the formalism is valid up to momentum of around $0.5-1$ GeV$^2$, \cite{Colangelo:2014dfa} will have to consider a matching with a certain model to account for the QCD high-energy behavior. How such matching should be done, and how model dependent would that be is not yet discussed. And how that would improve on previous results has still to be seen. At least, is a clean way to use experimental information of the $\pi^0$ TFF at low energies. We do not have such low-energy space-like data yet (despite significant constraints provided by time-like input, see~\cite{Hoferichter:2014vra}) but KLOE, MAMI, and BES are facilities where it could be in principle measured. More interesting would be the $2\pi$ contribution, since less is known from model calculations (see the third column on Table~\ref{T2}). Here the complication arises from the nature of the $\gamma^* \gamma^* \to \pi \pi$ process which, in DR formalism, is pretty involved. Without many assumptions, one expects a reliable contribution up to $0.5$ GeV$^2$ momentum transfer, i.e., only $S$-wave contribution. The inclusion of a $D$ wave makes the system of coupled DR much more complicated and less user friendly. On top, kaon and multi-pion contributions are neglected, and such assumption is difficult to quantify.

Beyond holographic QCD~\cite{Hong:2009zw,Cappiello:2010uy} or non-local quark model~\cite{Dorokhov:2011zf} and Dyson Schwinger~\cite{Eichmann:2014ooa} approaches, the common attempt to calculate the HLBL is through hadronic models constrained by data. In this framework it is easier to show the role of experimental data, specially for the pseudoscalar exchange contribution, which is driven by its TFF. 

The main obstacle when using experimental data is the lack of them, specially on the doubly virtual TFF. Fortunately, data on the TFF when one of the photons is real is available from different collaborations, not only for $\pi^0$ but also for $\eta$ and $\eta'$. It is common to factorize the TFF, i.e., $F_{P\gamma^*\gamma^*}(Q_1^2,Q_2^2) = F_{P\gamma^*\gamma}(Q_1^2,0) \times F_{P\gamma \gamma^*}(0,Q_2^2)$, and describe it based on a rational function. One includes a modification of its numerator due to the high-energy QCD constraints. Although the high-energy region of the model is not very important, it still contributes around a $20\%$. More important is the double virtuality, specially if one uses the same model (as it should) for predicting the $\pi^0 \to e^+e^-$ decay. Current models cannot accommodate its experimental value \cite{Benayoun:2014tra,Masjuan:2015lca}. The worrisome fact is that modifying the model parameters to match such decay and going back to the HLBL, would result in a dramatic decrease of the value~\cite{Masjuan:2015lca} and this is not properly discussed yet.

The HLBL is dominated by $Q^2$ from $0$ to $2$ GeV$^2$, and picked up around $0.5$ GeV$^2$. Then, good description of TFF in such region is very important. Such data are not yet available, but any model should reproduce the available one. That is why the authors of \cite{Masjuan:2012wy,Masjuan:2012qn,Escribano:2013kba}, in contrast to other approaches, did not used data directly but the low-energy parameters (LEP) of the Taylor expansion for the TFF and reconstructed it \textit{via} PAs. The LECs certainly know about all the data at all energies and as such incorporates all our experimental knowledge at once. This procedure implies a model-independent result together with a well-defined way to ascribe a systematic error. In other words, this is the first procedure that can be considered an \emph{approximation}, in contrast to the \emph{assumptions} considered in other approaches. The LECs were obtained in~\cite{Masjuan:2012wy} for the $\pi^0$ and in~\cite{Escribano:2013kba,Escribano:2015nra} for $\eta$ and $\eta'$. The HLBL value from such approach is quoted under EMS in Table~\ref{T2}.
I want to emphasize the role of experimental data. Let me use the LECs together with the $\pi^0 \to \gamma \gamma$ to match the free parameters of the LMD+V model introduced in~\cite{Knecht:2001qf}, as resonances or $F_{\pi}$. I obtain $a_{\mu}^{\mathrm{HLBL},\pi^0}=7.5\times 10^{-10}$ which contrasts with the original $a_{\mu}^{\mathrm{HLBL},\pi^0}=6.3\times 10^{-10}$~\cite{Knecht:2001qf}. The role of the new experimental data is then clear, inducing a $20\%$ effect. On top, since the LMD+V is not a PA but a well-educated model, it is difficult to ascribe a systematic error due to the large-$N_c$ approach. PAs already consider such corrections and Refs.~\cite{Masjuan:2012wy,Masjuan:2012qn,Escribano:2013kba} found them to be of the order of $5\% - 10\%$ less dramatic than the naive $30\%$ from the $N_c$ counting, but still to be taken into account.

The last word is about a complete different approach based on the Laporta and Remiddi (LR)~\cite{Laporta:1992pa} analytical result for the heavy quark contribution to the LBL. The idea is to extend the perturbative result to hadronic scales low enough for accounting at once for the whole HLBL. The free parameter is the quark mass $m_q$. The recent estimates using such methodology~\cite{Pivovarov:2001mw,Erler:2006vu,Boughezal:2011vw,Greynat:2012ww} found $m_q \sim 0.150 - 0.250$ GeV after comparing the particular model with the HVP. The value for the HLBL is higher than those shown in Table~\ref{T2}, around $a_{\mu}^{\mathrm{HLBL}}=(12 - 16)\times 10^{-10}$, which seems to indicate that the missing pieces of the standard calculations are not so negligible. Recently, \cite{Masjuan:2012qn} considered a different approach to the HLBL based also on the quark model. The LR model for the HLBL does not reproduce the well-known QCD features of symmetry breaking at low-energies (i.e., instead of a $\log (\Lambda/m_{\mu})^2$ divergency with $\Lambda$ a hadronic scale, the LR vanishes as $m_{\mu}^2/(2m_q)$ for large $m_q$, being $\Lambda \sim 2 m_q$.) QCD tells us that the pion contribution and the quark loop would decouple for large $\Lambda$. Since the hadronic scale turns out to be  $\sim M_{\rho}$ but not infinity, \cite{Masjuan:2012qn} searched for the scale where both contribution would coincide through averaging the photon momenta, finding $m_q \sim 170$ GeV. As such, they could also predict the HVP with great accuracy. They also found higher HLBL value, $a_{\mu}^{\mathrm{HLBL}}=15(2)\times 10^{-10}$.

Concluding, the reference numbers~\cite{Jegerlehner:2009ry,Prades:2009tw} seem robust and the QCD features of the HLBL seem to be well understood but there is \emph{a but}. The new experimental data seem to reveal larger contributions from pseudsocalar pieces, meaning that the modeling of the TFF is more important than expected. Also, systematic errors due to both chiral and large-$N_c$ limits are important and difficult to evaluate, but PAs can help. Both lattice QCD and DR are promising, but both suffer from different drawbacks. On top, the ballpark predictions coincide on drawing scenarios with larger values, indicating in my opinion the need to better understand the process. I think there is still a long way to go.

 \section*{Acknowledgements}
I thank the organizers for the very nice atmosphere during the conference and also S. Peris, M. Vanderhaeghen, E. Ruiz-Arriola, W. Broniowski, R. Escribano, S. Scherer, P. Roig and P. Sanchez-Puertas for discussions. Supported by the Deutsche Forschungsgemeinschaft through the CRC 1044.


\bibliographystyle{elsarticle-num}







\end{document}